\documentclass[11pt]{article}

\usepackage{amssymb}
\usepackage[dvips]{graphicx}
\usepackage{epsfig}
\usepackage{cite}

\def\laq{~\raise 0.4ex\hbox{$<$}\kern -0.8em\lower 0.62
ex\hbox{$\sim$}~}
\def\gaq{~\raise 0.4ex\hbox{$>$}\kern -0.7em\lower 0.62
ex\hbox{$\sim$}~}

\def\be{\begin{equation}}
\def\ee{\end{equation}}
\def\bea{\begin{eqnarray}}
\def\eea{\end{eqnarray}}
\newcommand{\nn}{\nonumber}
\newcommand{\de}{\partial}

\def \ra {\rightarrow}
\def \l {\lambda}
\def \L {\Lambda}

\def \g {\gamma}
\def \s {\sigma}

\def \e {\epsilon}

\def \Om {\Omega}

\def \5 {{}^{(5)}}
\def \cd {\nabla}

\begin{document}

\begin{flushright}

\end{flushright}

\vspace*{1.5 cm}

\begin{center}

\Huge
{Non-singular Brane cosmology with a Kalb-Ramond field}

\vspace{1cm}

\large{G. De Risi$^{1,2}$}

\vspace{.2in}

\normalsize
{\sl $^1$Institute of Cosmology \& Gravitation,
University of Portsmouth, Portsmouth~PO1 2EG, UK}\\

\vspace{.1in}

{\sl $^2$Istituto Nazionale di Fisica Nucleare, 00186~Roma, Italy}

\vspace*{1.5cm}

\begin{abstract}

We present a model in which a 3-brane is embedded in a warped 5-dimensional background with a dilaton and a Kalb-Ramond 2-form.
We show that it is possible to find static solutions of the form of charged dS/AdS-like black hole
which could have a negative mass parameter. The motion of the 3-brane in this bulk generates an effective
4-dimensional bouncing cosmology induced by the negative dark radiation term. This model avoids the instabilities
that arises for previous non-singular braneworld cosmologies in a Reissner-Nordstr{\o}m-AdS bulk.

\end{abstract}
\end{center}


\section{Introduction}

Braneworld models \cite{Randall:1999ee,Randall:1999vf} have generated, during the past decade,
enormous attention, due to the dramatic change they
inspired in our understanding of extra dimensions. According to this framework, our universe is a ``brane''
embedded in a higher-dimensional space, on which the Standard Model fields are confined, while gravity
is localised near the brane by the warped geometry of the extra dimension. It is possible to
construct models in which the brane evolution mimics a Friedmann-Robertson-Walker (FRW) cosmology,
with modifications at small scales due to the gravitational effect of the bulk spacetime on the brane \cite{Binetruy:1999ut,Binetruy:1999hy,Shiromizu:1999wj}.
In particular, provided the bulk is taken to be a Reissner-Nordstr{\o}m-AdS black hole, such modifications
can lead to bouncing 4D cosmological models \cite{Mukherji:2002ft}. Unfortunately the brane, during its evolution
in the bulk, always crosses the Cauchy horizon of the AdS black hole, which is unstable \cite{Kanti:2003bx,Hovdebo:2003ug}.

In this paper we present a different model \cite{DeRisi:2007dn}, in which this problem is avoided. We consider
a brane embedded in a supergravity background in which both the dilaton and the Kalb-Ramond 2-form are
turned on (but without a dilaton potential). By dualizing the 2-form, we obtain Einstein-Maxwell like equations
of motion, but with a different sign for the kinetic term of the Maxwell-like field. The static solution is therefore
different, and the term that dominates at high curvature is like ``stiff matter'' with positive energy
density. Even though this implies that the energy contribution at high curvatures is positive, so that
it can not drive a bounce, this opens an interesting possibility of having negative energy contributions
at intermediate curvatures, by letting the mass of the black hole be negative.
The parameter space allows this while avoiding a naked singularity. In this case, we show that
it is possible that the brane bounces before crossing the black hole horizon, so that the effective
4-dimensional cosmological evolution will not suffer from any instability \cite{Hovdebo:2003ug}.

\section{Bulk solution}

We will consider the low-energy string effective action:
\bea
S &=& \frac{M^3}{2} \int d^5 x \sqrt{- g } \left( R - 2\L e^{\s_1 \phi} - \frac{1}{12}H_{ABC}H^{ABC}e^{\s_2 \phi}
-g^{AB} \de_A \phi \de_B \phi \right) \nn \\
& & -  T_3 \int d^{4} \xi \sqrt{-\g} e^{\l \phi}
\label{Action}
\eea
(with $H_{ABC}=\de_{[A}B_{BC]}$), which describes a 3-brane embedded in a 5-dimensional bulk with dilaton
and Kalb-Ramond 2-form. We will take into account the presence of the brane, which is assumed to be neutral with
respect to the antisymmetric field, by implementing the Israel junction condition in the next
section \cite{Kraus:1999it}. The equation of motions derived from (\ref{Action}) can be greatly simplified
if we take, for the Kalb-Ramond field, the ansatz
\be
H^{CAB} = \e^{CABDE}\cd_D A_E e^{-\s_2 \phi}.
\label{HAnsatz}
\ee
by which we get the folowing equations of motion:
\bea
&& G_{AB} = \left[ -e^{\s_1 \phi}\L - \frac{1}{2} \left( \de_c \phi \right)^2 + \frac{1}{2} e^{-\s_2 \phi} F^2
\right] g_{AB} + \de_A \phi \de_B \phi - 2 e^{-\s_2 \phi} F_{AC}F_B^{~~C},
\label{DualEinsteq} \\
&& \cd_A \cd^A \phi - \s_1  e^{\s_1 \phi}\L - \frac{\s_2}{2} e^{-\s_2 \phi} F^2 = 0,
\label{DualDileq} \\
&& \cd_B \left( e^{-\s_2 \phi}F^{AB}\right) = 0,
\label{DualFeq}
\eea
with $F_{MN} = \frac{1}{2} \left( \cd_M A_N - \cd_N A_M \right)$ being a ``pseudo'' Maxwell field strength dual to
$H_{ABC}$. A static solution with a maximally symmetric 3-space and purely electric field can be cast as:
\bea
ds^2 &=& -\left( -\frac{Q^2}{3R^4} - \frac{\mu}{R^2} + k - \frac{\L}{6} R^2 \right) dt^2 + \frac{dR^2}{\left( -\frac{Q^2}{3R^4} - \frac{\mu}{R^2} + k - \frac{\L}{6} R^2 \right)} \nn \\
&& + R^2 \left( \frac{dr^2}{1-k r^2} + r^2 d\Om^2 \right) \nn \\
A(R) &=& \pm \frac{Q}{R^2}.
\label{backgroundsol}
\eea

In this solution it is possible to set $\mu<0$, which means that the mass of the central body is negative,
without having a naked singularity. In fact, assuming that the bulk cosmological constant is negative, $\L<0$,
we have an horizon located at
\be
R_0 = \left(\frac{2\mu}{\L}\right)^{1/4} \sqrt{{\rm C_{1/3}} \left( \frac{2Q^2\sqrt{-\L}}{(-2\mu)^{3/2}}\right)},
\label{singlehor}
\ee
where $C_{1/3}$ is the Chebyshev cubic root \cite{DeRisi:2007dn}. In the next section we will describe how
this bulk affects the cosmology of an embedded moving brane.

\section{Cosmology on the Brane}

\begin{figure}[ht]
\begin{center}
\epsfig{file=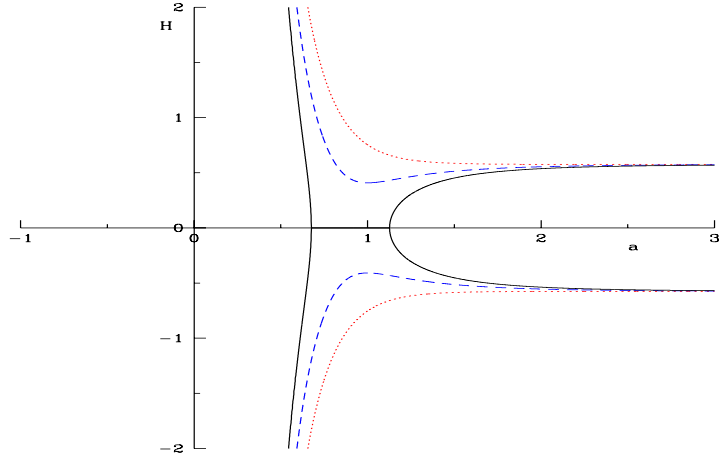,width=11cm,height=7cm}
\caption{The Hubble parameter $H$ as a function of the scale factor $a$. The three colours represent decreasing
values of $\frac{R_{KR}}{\ell_4}$: $15.81$ (red, dotted), $1.41$ (blue, dashed), $0.70$ (black, solid).}
\label{H_vs_a}
\end{center}
\end{figure}

The movement of an embedded brane in the 5D bulk induces a cosmological evolution on the brane
via the Israel junction condition \cite{Kraus:1999it}. In the case under investigation, if we assume
for simplicity a pure tension, spatially flat brane, the modified Friedmann equation is
\be
H^2 = \frac{\L_4}{3} + \frac{\mu}{a^4} + \frac{Q^2}{3a^6}.
\label{Fried}
\ee
The behaviour of $H$ as a function of $a$ is depicted in Fig. \ref{H_vs_a}
for different values of the ratio between the 4D de Sitter curvature radius and the characteristic length of the
Kalb-Ramond black hole obtained by the charge to mass ratio $R_{KR} = \frac{Q^2}{2(-\mu)^{3/2}}$.

The black plot have two values of $a$ for which $H = 0$. It is not difficult to understand that the largest branch of this
plot describes a bouncing universe that approaches asymptotically to de Sitter The bounce occurs at the following
value of the scale factor:
\be
a_b = \sqrt{2} \left(-\frac{\mu}{\L_4}\right)^{1/4} \sqrt{\cos \left[ \frac{1}{3}
\arccos \left( -\frac{Q^2\sqrt{\L_4}}{2(-\mu)^{3/2}}\right)\right]}.
\label{abounce}
\ee

\begin{figure}[ht]
\begin{center}
\epsfig{file=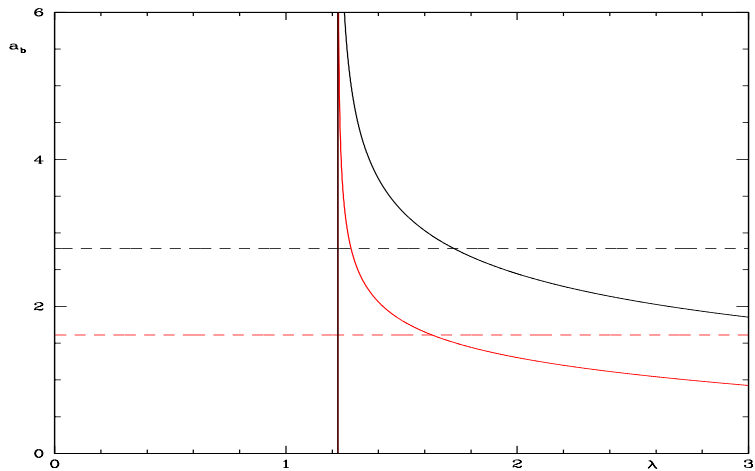,width=11cm,height=7cm}
\caption{Plot of $a_b$ as a function of the tension $\l$ for different values of $\frac{R_{KR}}{\ell_5}$: $0.5$
(black, darker), $0.02$ (red, lighter). Dashed curves of the same colour represent the position of the horizons for the same value of $\frac{R_{KR}}{\ell_5}$.}
\label{horcomp}
\end{center}
\end{figure}

Now we have to show that the bounce occurs outside the horizon. Fig. \ref{horcomp} shows $a_b(\l)$, and the corresponding
value of $R_0$. We can see that there is a region, when the brane tension is
close to the minimum allowed value, in which the bounce radius is greater than the horizon position,
so that the entire evolution of the brane lies in the physically viable region outside the horizon.
This feature is quite general, and the reason is easy to understand, since we can see from
(\ref{abounce}) that $a_b \ra \infty$ as $\L_4 \ra 0$.

Analytically, we find that $\L_4$ has to satisfy the following inequalities:
\bea
\L_4 < -\frac{\L}{2}\frac{\frac{3}{2}C_{1/3}\left(2\frac{Q^2\sqrt{-\L}}{(-2\mu)^{3/2}}\right) -
\frac{Q^2\sqrt{-\L}}{(-2\mu)^{3/2}}}
{\frac{3}{2}C_{1/3}\left(2\frac{Q^2\sqrt{-\L}}{(-2\mu)^{3/2}}\right) + \frac{Q^2\sqrt{-\L}}{(-2\mu)^{3/2}}}
&~{\rm for}~& \frac{Q^2\sqrt{-\L}}{(-2\mu)^{3/2}}<5, \\
\L_4 < -\frac{\L}{4} C_{1/3}^{-2}(2\frac{Q^2\sqrt{-\L}}{(-2\mu)^{3/2}}) &~{\rm for}~& \frac{Q^2\sqrt{-\L}}{(-2\mu)^{3/2}}>5,
\label{4DLint}
\eea
So there is always an allowed value of the brane tension for which the brane evolution lies entirely outside the horizon,
and therefore free of instabilities.

\section{Conclusions and outlook}

In this paper we presented a braneworld model in which the cosmological evolution of the brane is non-singular, and
the brane universe bounces smoothly from a phase of contraction to a subsequent expanding phase.
The cosmological evolution on the brane is induced by its movement through a static bulk AdS black
hole supported by a non-trivial Kalb-Ramond antisymmetric 2-form. The bouncing is driven by the negative dark
radiation that appears as a peculiar feature of the bulk solution, which is a negative mass black hole.
Further investigations would be needed to clarify the main issues that may arise in developing this model, namely the
overall stability of the 5D negative mass black hole solution and the inclusion of radiation on the brane,
which could spoil the singular-free behaviour. Other interesting development would be the study of the model in presence
of a DBI coupling between the brane and the Kalb-Ramond field, which arises naturally in the context of
String Theory, and the analysis of perturbations, in order to compare the model to observations.

\section*{Acknowledgements}
I would like to thank the I.N.F.N. for providing fundings to attend the ``$43^{rd}$ Rencontres de Moriond'' conference, where
this talk was given.

\end{document}